\documentclass[%
 aip,
 rsi,
amsmath,amssymb,
reprint,
floatfix
]{revtex4-1}

\usepackage{graphicx}
\usepackage{dcolumn}
\usepackage{bm}

\usepackage[utf8]{inputenc}
\usepackage{mathptmx}
\usepackage{braket}
\usepackage[english]{babel}
\usepackage{adjustbox}
\usepackage{xcolor}
\colorlet{RED}{red}
\colorlet{BLUE}{blue}
\usepackage{dcolumn}
\usepackage{bm}
\usepackage[version=4]{mhchem} 
\usepackage{acronym}
\usepackage{adjustbox}
\usepackage{tikz}
\usepackage{microtype} 
\usetikzlibrary{calc,shapes.geometric,decorations.pathmorphing,patterns}

\definecolor{background-color}{gray}{0.98}
\usepackage[margin=2.3cm,bmargin=1cm,footnotesep=1cm]{geometry}

\begin{document}

\title{Coupled Cluster Downfolding Methods: 
the effect of double commutator terms 
on the accuracy of ground-state energies}

\author{Nicholas P. Bauman}
\email{nicholas.bauman@pnnl.gov}
 \affiliation{Physical Sciences and Computational Division, Pacific Northwest National Laboratory, Richland, Washington 99354, United States of America}
\author{Karol Kowalski} 
\email{karol.kowalski@pnnl.gov}
 \affiliation{Physical Sciences and Computational Division, Pacific Northwest National Laboratory, Richland, Washington 99354, United States of America}
\date{April 2020}

\date{\today}

\begin{abstract}
Downfolding coupled cluster (CC) techniques have recently been introduced into quantum chemistry as a tool for the dimensionality reduction of the many-body quantum problem. As opposed to earlier formulations in physics and chemistry based on the concept of effective Hamiltonians, the appearance of the downfolded Hamiltonians is a natural consequence of the single-reference exponential parametrization of the wave function. In this paper, we discuss the impact of higher-order terms originating in double commutators. In analogy to previous studies, we consider the case when only one- and two-body interactions are included in the downfolded Hamiltonians. 
We demonstrate the efficiency of the many-body expansions involving single and double commutators for the unitary extension of the downfolded Hamiltonians on the example of the beryllium atom, and bond-breaking processes  in the  Li$_2$ and H$_2$O molecules. For the H$_2$O system, we also analyze energies obtained with downfolding procedures as functions of the active space size. 
\end{abstract}

\maketitle

\section{Introduction} 
 
Over the last few decades, the coupled cluster (CC) theory \cite{coester58_421,coester60_477,cizek66_4256,paldus72_50,purvis82_1910,jorgensen90_3333,paldus07,crawford2000introduction,bartlett_rmp} has 
evolved into one of the most accurate and dominant  theories to describe  various  many-body systems across spatial scales
and therefore addressing fundamental problems in 
quantum chemistry, 
\cite{scheiner1987analytic,sinnokrot2002estimates,slipchenko2002singlet,tajti2004heat,crawford2006ab,parkhill2009perfect,riplinger2013efficient,yuwono2020quantum}
and material sciences,
\cite{stoll1992correlation,hirata2004coupled,katagiri2005equation,booth2013towards,degroote2016polynomial,mcclain2017gaussian,wang2020excitons,PhysRevX.10.041043}
and 
nuclear structure theory, 
\cite{PhysRevC.69.054320,PhysRevLett.92.132501,PhysRevLett.101.092502}
Many  strengths of the single-reference CC formalism (SR-CC)
originate in the exponential representation  of the ground-state wave function $|\Psi\rangle$,
\begin{equation}
|\Psi\rangle = e^T |\Phi\rangle \;,
\label{ccwf}
\end{equation}
where $T$ and $|\Phi\rangle$ correspond to the cluster operator and a reference function that provides an approximation to the exact ground-state wave function. For practical applications, one can define a hierarchy of CC approximations by increasing the excitation level in the cluster operator. Another important feature of the CC formalism stems from the linked cluster theorem \cite{brandow67_771,lindgren12} which allows one to build efficient algorithms for the inclusion of the higher-rank excitations. 
When these two features are combined, they give rise to efficient and accurate methodologies that account for higher-order correlation effects and that have been widely used in physics and quantum chemistry. The diverse manifold of higher-order approximations include, among others, categories of active-space, multi-reference, externally corrected, and adaptive approaches, as well as non-iterative corrections and approaches that combine stochastic techniques (for key reviews and papers see Refs. (\onlinecite{bartlett_rmp, paldus07, Jeziorskirev, paldus2010multireference, EvangelistaMR, piecuch_molphys, bauman1, kowalski2018regularized, Paldus2017, thom2010stochastic,deustua2017converging}) and references therein).


Recently, it was shown using the sub-system embedding sub-algebras (SES) \cite{safkk} approach that the CC formalism is a natural renormalization procedure that allows one to construct effective Hamiltonians in reduced-dimensionality active spaces for classes of  sub-systems of the undivided quantum system. 
This formalism provides a mathematically rigorous procedure to construct the many-body form of the active-space effective Hamiltonians using only external (out of active space) Fermionic degrees of freedom that can be extracted from single reference CC wave function expansions.
In contrast, the internal degrees  of freedom (inside active space) are determined by the  diagonalization of effective Hamiltonians in the active space.
The SES-CC formalism is also valid for any sub-system, and the resulting sub-problems (or computational blocks) can be integrated into the computational flows (or quantum flow algorithms described in  Ref. \onlinecite{kowalski2021dimensionality}), which can traverse large sub-spaces of the corresponding Hilbert space. For example, using this approach, one can define flows involving  occupied localized orbitals, bypassing certain problems of traditional local CC formulations. Additionally, SES-CC can be extended to the time domain, leading to time-dependent Hamiltonians that describe the dynamics of the entire system in the active space (provided that time-dependent external amplitudes are known or can be efficiently approximated/evaluated). 

In the light of the above  discussion, high-accuracy  CC-based techniques for novel representations of quantum many-body problem in reduced-dimensionality spaces may play an important  role not only in solving complex problems but also in enabling new forms of simulations  associated with the emergence of the  quantum computing (QC) technologies. Although reduced-dimensionality methods have been existing for long time and are usually identified with Complete Active Space Self-Consistent Field (CASSCF) methods (see for example Refs. \onlinecite{olsen2011casscf,ma2011generalized} and references therein),  
density matrix renormalization group  (DMRG) methods,\cite{white1992density,schollwock2005density,dmrg2,chan2011density} 
local CC formulations, 
\cite{hampel1996local,schutz2000low,schutz2000local,schutz2001low,li2002linear,li2006efficient,li2009local,li2010multilevel,neese2009efficient,neese2009accurate,Neese16_024109,riplinger2013natural,pavosevic2016}, 
and correlated approaches where the energies are obtained through the diagonalization of the effective Hamiltonians in reduced-size sub-spaces (or equivalently - model spaces), \cite{bloch1958theorie,des1960extension,lowdin1963studies,schrieffer1966relation,brandow1967linked,soliverez1969effective,schucan1972effective,jorgensen1975effective,mukherjee1975correlation,jezmonk,kutzelnigg1982quantum,
stolarczyk1985coupled,
mukherjee1986linked,durand1983direct,
durand1987effective,jeziorski1989valence,kaldor1991fock,rittby1991multireference,andersson1990second,andersson1992second,
hirao1992multireference,finley1995applications,
glazek1993renormalization,meissner1995effective,
meissner1998fock,nakano1998analytic,
angeli2001n,angeli2001introduction,
mrcclyakh,bravyi2011schrieffer,sahinoglu2020hamiltonian}
the SES-CC formulation allows one to construct active-space effective Hamiltonins using single-reference concepts. A problem with SES-CC effective Hamiltonians in certain applications, such as quantum computing,  is their non-Hermitian character. In light of this problem, we have recently explored the double 
unitary coupled cluster Ansatz (DUCC), 
\cite{bauman2019downfolding,downfolding2020t,metcalf2020resource,bauman2019quantumex,bauman2020variational}
which, in analogy to the SES-CC techniques, results in an active-space representation of quantum problem with the added benefit that the resulting DUCC effective Hamiltonians are Hermitian.

In analogy to the canonical SES-CC approach and various effective Hamiltonian formulations based on the wave operator formalism and/or similarity transformations, the DUCC formalism also decouples the external and internal Fermionic degrees of freedom. 
These Hamiltonians have been recently studied in the context of quantum simulations using various quantum solvers, including Quantum Phase Estimator (QPE),\cite{luis1996optimum,
cleve1998quantum,berry2007efficient,childs2010relationship,wecker2015progress,
haner2016high,poulin2017fast} and
Variational Quantum Eigensolvers (VQE),\cite{peruzzo2014variational,mcclean2016theory,romero2018strategies,PhysRevA.95.020501,Kandala2017,kandala2018extending,PhysRevX.8.011021,huggins2020non,ryabinkin2018qubit,cao2019quantum,ryabinkin2020iterative,izmaylov2019unitary,lang2020unitary,grimsley2019adaptive,grimsley2019trotterized} where representations of the DUCC effective Hamiltonians used in these studies were defined by the lowest-order contributions to the corresponding commutator expansion. The area of similarity transformations of electronic Hamiltonians induced by exponential operators involving cluster operators expressed in terms of non-commutative algebras is under active development 
(for recent developments see Refs. \onlinecite{evangelista2011a,li2020connected,evangelista2019exact}).

In this manuscript, we will extend this analysis to DUCC effective Hamiltonians that include higher-order correlations effects. In particular, we will focus on approximations involving double commutators. 
The main focus will be on  the approximate schemes where the class of external excitations used to define DUCC effective Hamiltonians includes singly and doubly excited amplitudes that are used to construct one- and two-body effective interactions. Within this approximation, we consider algorithms that are correct up to the third order of the standard M\o ller--Plesset perturbation theory. We illustrate the performance of these methods on the example of several benchmark systems used in recent studies of 
quantum algorithms based on the downfolding formalisms. This includes the Be,  Li$_2$, and H$_2$O systems.
Additionally, the Li$_2$ system epitomizes the situation where the CCSD level of theory provides an accurate  approximation to the exact (full configuration interaction (FCI)) formalism, and the active space can be properly defined. Using this example,  we demonstrate that in this situation, the approximations defined by double commutators can provide a systematic improvement in the accuracy of energies obtained by the diagonalization of the  effective Hamiltonians defined by single-commutator-based approaches. For the  H$_2$O system, we discuss the behavior of the DUCC energies as functions of active space size. These results provide yet another illustration of the feasibility of the CC downfolding techniques.

\section{The DUCC effective Hamiltonian  formalism}

The DUCC effective Hamiltonian formalism has been amply discussed in recent papers.\cite{bauman2019downfolding,bauman2019quantumex,downfolding2020t,kowalski2021dimensionality} Here we only describe basic tenets of this approach. 
The DUCC formalism utilizes a composite 
unitary CC Ansatz  to represent the exact wave function $|\Psi\rangle$, i.e.,
\begin{equation}
        |\Psi\rangle=e^{\sigma_{\rm ext}} e^{\sigma_{\rm int}}|\Phi\rangle \;,
\label{ducc1}
\end{equation}
where $\sigma_{\rm ext}$ and $\sigma_{\rm int}$ are general-type anti-Hermitian operators 
\begin{eqnarray}
\sigma_{\rm int}^{\dagger} &=&  -\sigma_{\rm int} \;, \label{sintah} \\
\sigma_{\rm ext}^{\dagger} &=&  -\sigma_{\rm ext} \;. \label{sintah2}
\end{eqnarray} 
All cluster amplitudes defining the $\sigma_{\rm int}$ cluster operator carry active  spin-orbital indices only and define all possible excitations within the corresponding active space. The external cluster operator  $\sigma_{\rm ext}$ is defined by amplitudes carrying at least one inactive spin-orbital index
(for more detaisl see the discussion of DUCC expansion in the Ref.\onlinecite{downfolding2020t}). 

Using the DUCC representation (Eq.(\ref{ducc1})), it can be shown that the energy of the entire system (once the exact form of $\sigma_{\rm ext}$ operator is known) can be calculated through the diagonalization of the effective/downfolded Hamiltonian in the active space, i.e., 
\begin{equation}
        H^{\rm eff} e^{\sigma_{\rm int}} |\Phi\rangle = E e^{\sigma_{\rm int}}|\Phi\rangle,
\label{duccstep2}
\end{equation}
where
\begin{equation}
        H^{\rm eff} = (P+Q_{\rm int}) \bar{H}_{\rm ext} (P+Q_{\rm int})
\label{equivducc}
\end{equation}
and 
\begin{equation}
        \bar{H}_{\rm ext} =e^{-\sigma_{\rm ext}}H e^{\sigma_{\rm ext}}.
\label{duccexth}
\end{equation}
In Eq.(\ref{equivducc}), $P$ and $Q_{\rm int}$ are the projection operators on the reference function $|\Phi\rangle$ and all excited Slater determinants (defined with respect to the $|\Phi\rangle$ reference) in the active space, respectively. 
Once the external cluster amplitudes are known (or can be effectively approximated),  the energy of the system  can be evaluated by diagonalizing Hermitian effective/downfolded Hamiltonian in the active space using various quantum/conventional-computing solvers. 
Important steps towards developing practical computational schemes based on the utilizaton of the DUCC  downfolded Hamiltonians are (1)  efficient approximation of the $\sigma_{\rm ext}$ operator, and (2)  approximate form of the   non-terminating commutator expansions defining downfolded Hamiltonians. 
A legitimate approximation for $\sigma_{\rm ext}$  is through the utilization of the external part of the standard cluster operator
\begin{eqnarray}
\sigma_{\rm ext} &\simeq& T_{\rm ext} - T_{\rm ext}^{\dagger} \;, \label{sext}
\end{eqnarray}
which has been discussed in Ref. \onlinecite{bauman2019downfolding}. 
In particular, $T_{\rm ext}$ can be approximated by SR-CCSD amplitudes that carry at least one external spin-orbital index. Other possible sources for evaluating external cluster amplitudes are higher-rank single-reference CC methods, truncated expansions,\cite{evangelista2011a,canonical1} and approximate unitary CC formulations such as UCC({\it n}) methods. \cite{unitary1,unitary2}

Given the fact that the $\bar{H}_{\rm ext}$ operator is defined as an infinite expansion in terms of the $\sigma_{\rm ext}$ operator, we cannot calculate its exact form.
 For practical reasons, in this paper, we implement a three-step  simplification procedure. The first step, denoted as a simplification S1, is the estimation of $\sigma_{\rm ext}$ through the canonical CCSD external amplitudes, i.e., $\sigma_{\rm ext}\simeq T_{\rm ext}^{(\rm CCSD)} - T_{\rm ext}^{(\rm CCSD) \dagger}$. This  simplification is justifiable as the CCSD approach is capable of capturing a large part of dynamical correlation effects, especially when there is a clear demarcation between active spaces contributing to static ($\sigma_{\rm int}$) and dynamical ($\sigma_{\rm ext}$) correlation. The second approximation, denoted S2, is to limit $\bar{H}_{\rm ext}$ to one- and two-body contributions. The last simplification, which we will explore in more detail in this work, is the truncation of the commutator expansion of $\bar{H}_{\rm ext}$ (denoted as S3).

In this work, in addition to the single commutator expansion used in the previous analysis of the CC downfolding \cite{bauman2019downfolding} we include non-trivial contributions stemming from the double commutators. We collected all the approximate versions of the downfolded Hamiltonians in Table \ref{table_approx}, where we analyze seven approximate schemes A(1)-A(7).  In the expressions for A(1)-A(7), 
\begin{equation}
        H=\sum_{p,q} h^p_q a_p^{\dagger} a_q + \frac{1}{4}\sum_{p,q,r,s} v^{pq}_{rs} 
    a_p^{\dagger} a_q^{\dagger} a_s a_r
\label{ham}
\end{equation}
and 
\begin{equation}
        H_{N}= H-\langle\Phi|H|\Phi\rangle \;,
\label{normham}
\end{equation}
where $p$,  $q$, $r$, $s$ are spin-orbital labels, $h^p_q$ and $v^{pq}_{rs}$ represent one- and antisymmetric two-electron integrals, and 
\begin{equation}
        H_{N}= F_{N} + V_{N} \;,
\label{normham2}
\end{equation}
where $F_{N}$ and $V_{N}$ are are one- and two-body components of the $H_{N}$ operator. The commutators terms in Table \ref{table_approx} are in the particle-hole normal product form and we refer the reader to Ref. \onlinecite{bauman2019downfolding} for the procedure on how to obtain the corresponding Hamiltonian in physical-vacuum normal product form.
The first approximation, A(1), utilizes bare Hamiltonian in the active space while
approximations A(2), A(3), and A(4) are driven by single commutators (in approximations  A(2) and A(4), $F_N$-dependent double commutator terms are added to provide perturbative consistency to the second order of many-body perturbation theory (MBPT)). Approximations A(5), A(6), and A(7) introduce non-trivial terms due to the second commutators. In approximations A(5) and A(7), in analogy to A(2) and A(4), $F_N$-dependent triple commutators are introduced to provide perturbative consistency at the third order of perturbation theory (MBPT(3)). 
\begin{widetext}
\renewcommand{\tabcolsep}{0.2cm}
\begin{center}
  \begin{table*}
    \centering
    \caption{Approximate forms of the $\bar{H}_{\rm ext}$ operator considered in this paper. Special notation is used to designate perturbative structure of expansions employed. For example, $X^{(2)}$ and $X^{(3)}$ designate parts of the operator $X$ correct up to the second and third order of MBPT. For the case of the natural orbitals the same expressions are used with the full non-diagonal form of the Fock operator.}
    \begin{tabular}{lc} \hline \hline  \\
A(1) & $\bar{H}_{\rm ext}^{A(1)} = H$ \\[0.2cm]
A(2) & $\bar{H}_{\rm ext}^{A(2)} = H+[H_N,\sigma_{\rm ext}]^{(2)} + \frac{1}{2}[[F_N,\sigma_{\rm ext}],\sigma_{\rm ext}]^{(2)}$ \\[0.2cm]
A(3) & $\bar{H}_{\rm ext}^{A(3)} = H+[H_N,\sigma_{\rm ext}] $ \\[0.2cm]
A(4) & $\bar{H}_{\rm ext}^{A(4)} =H+[H_N,\sigma_{\rm ext}] +\frac{1}{2}[[F_N,\sigma_{\rm ext}],\sigma_{\rm ext}] $ \\[0.2cm]
A(5) & $\bar{H}_{\rm ext}^{A(5)} = H+[H_N,\sigma_{\rm ext}]^{(3)} +\frac{1}{2}[[H_N,\sigma_{\rm ext}],\sigma_{\rm ext}]^{(3)} +\frac{1}{6} [[[F_N,\sigma_{\rm ext}],\sigma_{\rm ext}],\sigma_{\rm ext}]^{(3)}$ \\[0.3cm]
A(6) & $\bar{H}_{\rm ext}^{A(6)} = H+[H_N,\sigma_{\rm ext}]+\frac{1}{2}[[H_N,\sigma_{\rm ext}],\sigma_{\rm ext}]$ \\[0,3cm]
A(7) & $\bar{H}_{\rm ext}^{A(7)} = H+[H_N,\sigma_{\rm ext}]+\frac{1}{2}[[H_N,\sigma_{\rm ext}],\sigma_{\rm ext}]+\frac{1}{6} [[[F_N,\sigma_{\rm ext}],\sigma_{\rm ext}],\sigma_{\rm ext}]$ \\[0.3cm]	
    \hline \hline 
    \end{tabular}
    \label{table_approx}
  \end{table*}
\end{center}   
\end{widetext}
The perturbative analysis of the downfolded Hamiltonians involves the perturbative expansion for the $T_{\rm ext}$ operator. It should be, however, remembered that due to the construction of the active space
and the definition of the external part of the cluster operator, the riskiest amplitudes characterized by large values are not explicitly present in  $T_{\rm ext}$. Therefore, the perturbative analysis with $T_{\rm ext}$ may not involve typical problems encountered by the full $T$ operator for the challenging cases defined by the presence of strong correlation effects.

The efficiency of commutator expansion also hinges upon the judicious choice of the active space. When 
the active space adequately captures the static correlation effects leading to small values of  $\sigma_{\rm ext}$ amplitudes; the commutator expansion should provide a hierarchical 
class of approximations, 
otherwise, infinite commutator expansion involving large values of external amplitudes may lead to
divergent nature of expansions based on this scheme.

\section{Results and Discussion }

The implementations of A(2)-A(7) formulations have been integrated with the Tensor Contraction Engine (TCE)\cite{hirata03_9887} environment of NWChem,\cite{valiev10_1477,apra2020nwchem} which allows us to use a variety of formulations to extract external cluster operator. In the present studies, we use the TCE CCSD module to provide singly and doubly excited parts of the external cluster operator ($T_{\rm ext}$). 
Calculations were performed using restricted Hartree-Fock (RHF) and natural orbitals at the MBPT(2) level (for the Li$_2$ system only). 
Additionally, in all calculations, all occupied spin orbitals were considered active. 
%
%

\subsection{The beryllium atom}

\begin{figure}
	\includegraphics[trim=0 2.5cm 0 0, angle=0, width=0.36\textwidth]{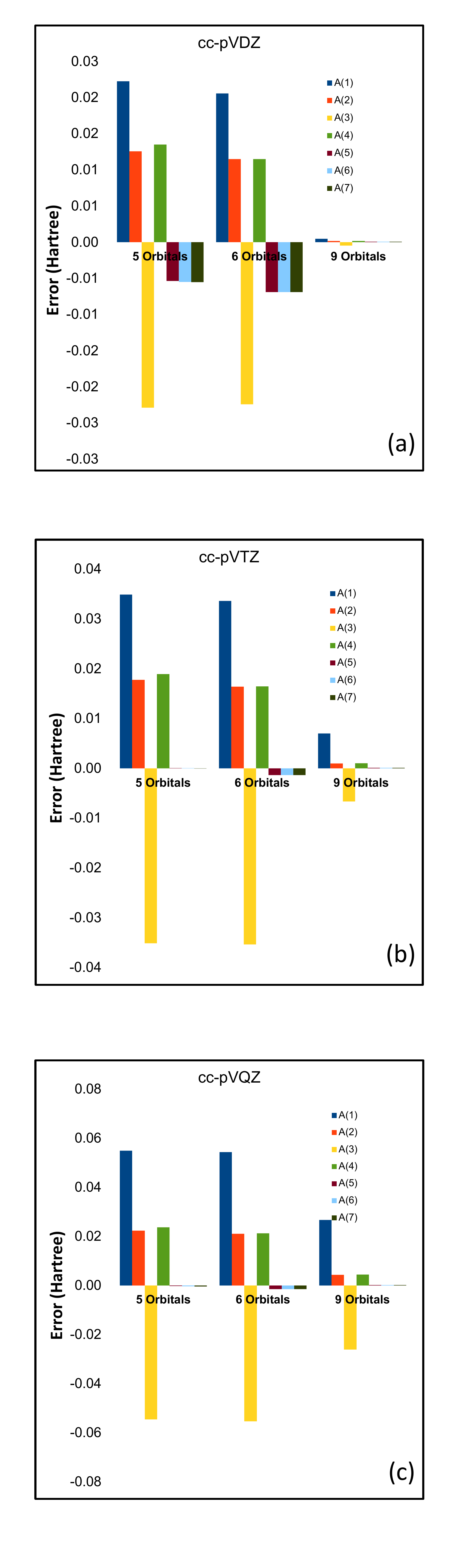}
	\caption{Comparison of energies (reported as errors with respect to the FCI energies)  obtained with A(1)-A(7) approximations in the cc-pVDZ (a), cc-pVTZ (b), and cc-pVQZ (c) basis sets. The horizontal axes represent a total  number of active  orbitals. In all calculations, RHF orbitals were used and all occupied orbitals were considered as active. 
    The  cc-pVDZ, cc-pVTZ, and cc-pVQZ generate 14, 30, and 55 orbitals, respectively.}
\label{fig1}
\end{figure}

The calculations for the beryllium atom were performed  using cc-pVDZ, cc-pVTZ, and cc-pVQZ basis sets in active spaces composed of 5, 6, and 9 lowest-lying RHF orbitals. 
The results of our simulations are shown in Fig. \ref{fig1} and Tables \ref{table_be1} and \ref{table_be2}. In Fig. \ref{fig1} we show the errors of  approximations A(1)-A(7) obtained with various active spaces with respect to the FCI results obtained when correlating all RHF orbitals in cc-pVXZ (X=D,T,Q) basis sets. One can observe a consistent pattern emerging: (1) the behavior of downfolded methods is similar for all basis sets employed, (2) the accuracy of the energies obtained with the effective Hamiltonians increase as the size of the active space increases and as higher-order commutators are included in the effective Hamiltonian expansion.
These observations are  corroborated by the results in Table \ref{table_be1}, where 
total energies are compared and Table \ref{table_be2}, where the percentage of total correlations energy reproduced by a given downfolded Hamiltonian is shown. 
The importance of the higher-order terms stemming from double commutator is well illustrated by the A(5)-A(7) methods for small (5 orbitals) active space in cc-pVQZ basis set (Table \ref{table_be2}) where these approaches can reproduce exact correlation energy. For this specific example, the performance of the single-commutator-driven approaches A(2)-A(4) is significantly worse. This is associated with the quality of the active space, which is reflected by the fact that diagonalizing bare Hamiltonian in the active space defined by 5 orbitals reproduces only 18 \% of the total ground-state correlation energy. 
Using larger active space for the cc-pVQZ basis set, defined by 9 lowest-lying RHF orbitals, results in single-commutator-driven approaches A(2) and A(4) performing much better than in the case of small 5-orbital active space. As discussed in Ref. \onlinecite{bauman2019downfolding,metcalf2020resource}, approximation A(3) is not fully consistent at the second order of  perturbation theory and may result in doubling the external correlation effects, which is reflected by the results shown in Fig. \ref{fig1}. One should also
observe systematic improvements of the quality of the A(2) and A(4) formulations with the size of the active space employed, which is a consequence of the inclusion of important $F_N$ dependent double commutator terms. 
For larger active spaces, double-commutator-driven approximations A(5)-A(7) approach FCI level of accuracy. 
The utilization of the larger active spaces is also advantageous from the point of minimizing the effects of additional approximations such as neglecting three-body interactions in the effective Hamiltonians  or using an approximate form of the $\sigma_{\rm ext}$ operator.
\renewcommand{\tabcolsep}{0.2cm}
\begin{center}
  \begin{table*}
    \centering
    \caption{Comparison of energies of the beryllium atom obtained with A(1)-A(7) approximations in the cc-pVDZ (a), cc-pVTZ (b), and cc-pVQZ (c) basis sets. In all calculations, RHF orbitals were used and all occupied orbitals were considered as active. The A(1) energies for all orbitals correlated correspond to the FCI energy of the system.
    The  cc-pVDZ, cc-pVTZ, and cc-pVQZ basis sets generate 14, 30, and 55 orbitals, respectively. 
}
\begin{tabular}{lccccc} \hline \hline  \\
                  & \multicolumn{4}{c}{Size of active space} &  \\
Method & 5 orbitals & 6 orbitals & 9 orbitals & All orbitals  &HF energy \\[0.1cm] 
    \hline \\[-0.2cm]
\multicolumn{6}{c}{cc-pVDZ} \\[0.2cm]
A(1) 	 & -14.59517	& -14.59683 &	-14.61692 &	-14.61741 &	-14.57234 \\
A(2) & -14.60483 &	-14.60590 &	-14.61723 & \\
A(3) & -14.64027 &	-14.63981 &	-14.61788 & \\
A(4) & -14.60390 &	-14.60590 &	-14.61723 & \\
A(5) & -14.62276 &	-14.62431 & -14.61732 & \\
A(6) & -14.62290 &	-14.62431 &	-14.61732 & \\
A(7) & -14.62293 &	-14.62431 &	-14.61732 & \\
\hline \\[-0.2cm]
\multicolumn{6}{c}{cc-pVDZ} \\[0.2cm]
A(1) & -14.58893 &	-14.59019&	-14.61679&	-14.62381 &	-14.57287 \\
A(2) & -14.606074 &	-14.607405 & 	-14.622826 &  \\
A(3) & -14.658945 &	-14.659147 &	-14.630455 & \\
A(4) & -14.604874 &	-14.607367 &	-14.622796 & \\
A(5) & -14.623735 &	-14.625154 &	-14.623715 &  \\
A(6) & -14.623786 &	-14.625158  &	-14.623714 & \\
A(7) & -14.623818 &	-14.625161 &	-14.623715 & \\
\hline \\[-0.2cm]
\multicolumn{6}{c}{cc-pVQZ} \\[0.2cm]
A(1)  &	-14.58520 &	-14.58578&	-14.61335&	-14.64012&	-14.57297\\
A(2) & -14.617794 &	-14.618964 & -14.635782 & \\
A(3) & -14.694742 &	-14.695527 & -14.666215 & \\
A(4) & -14.616380 &	-14.618790 & -14.635627 & \\
A(5) & -14.640296 &	-14.641596 & -14.639972 & \\
A(6) & -14.640455 &	-14.641553 & -14.639915 & \\
A(7) & -14.640494 &	-14.641560 & -14.639919 & \\[0.2cm]
    \hline \hline 
    \end{tabular}
    \label{table_be1}
  \end{table*}
\end{center}   
\renewcommand{\tabcolsep}{0.2cm}
\begin{center}
  \begin{table*}
    \centering
    \caption{Comparison of the percentage of correlation energy of the Be atom recovered by A(1)-A(7) approximation with respect to the  FCI correlation energy obtained in calculations correlating  all orbitals. }
    \begin{tabular}{lccc} \hline \hline  \\
                  & \multicolumn{3}{c}{Size of active space}  \\
Method & 5 orbitals & 6 orbitals & 9 orbitals  \\[0.1cm] 
    \hline \\[-0.2cm]
\multicolumn{4}{c}{cc-pVDZ} \\[0.2cm]
A(1) & 50.7	& 54.3 &	98.9 \\
A(2) & 72.1	& 74.5 &	99.6 \\
A(3) & 49.3	& 50.3 & 	99.0 \\
A(4) & 70.0	& 74.5 &	99.6 \\
A(5) & 88.1	& 84.7 &	99.8 \\
A(6) & 87.8	& 84.7 &	99.8 \\
A(7) & 87.8	& 84.7 &	99.8 \\
\hline \\[-0.2cm]
\multicolumn{4}{c}{cc-pVTZ} \\[0.2cm]
A(1) &31.5 &	34.0 &	86.2 \\
A(2) &65.2 &	67.8 &	98.1 \\
A(3) &31.0 &	30.6 &	87.0 \\
A(4) &62.8 &	67.7 &	98.0 \\
A(5) &99.9 &	97.4 &	99.8 \\
A(6) &100.0& 	97.4 &	99.8 \\
A(7) &100.0&	97.3 &	99.8 \\
\hline \\[-0.2cm]
\multicolumn{4}{c}{cc-pVQZ} \\[0.2cm]
A(1) & 18.2	& 19.1 &	60.1 \\
A(2) & 66.7	& 68.5 &	93.5 \\
A(3) & 18.7	& 17.5 &	61.1 \\
A(4) & 64.6	& 68.2 &	93.3 \\
A(5) & 99.7	& 97.8 &	99.8 \\
A(6) & 99.5	& 97.9 &	99.7 \\
A(7) & 99.4	& 97.9 &	99.7  \\[0.2cm]		
    \hline \hline 
    \end{tabular}
    \label{table_be2}
  \end{table*}
\end{center}   
%
%
%
%
%
%
%
%
%
%
%
\subsection{The Li$_2$ system}
\begin{figure}
	\includegraphics[angle=0, width=0.48\textwidth]{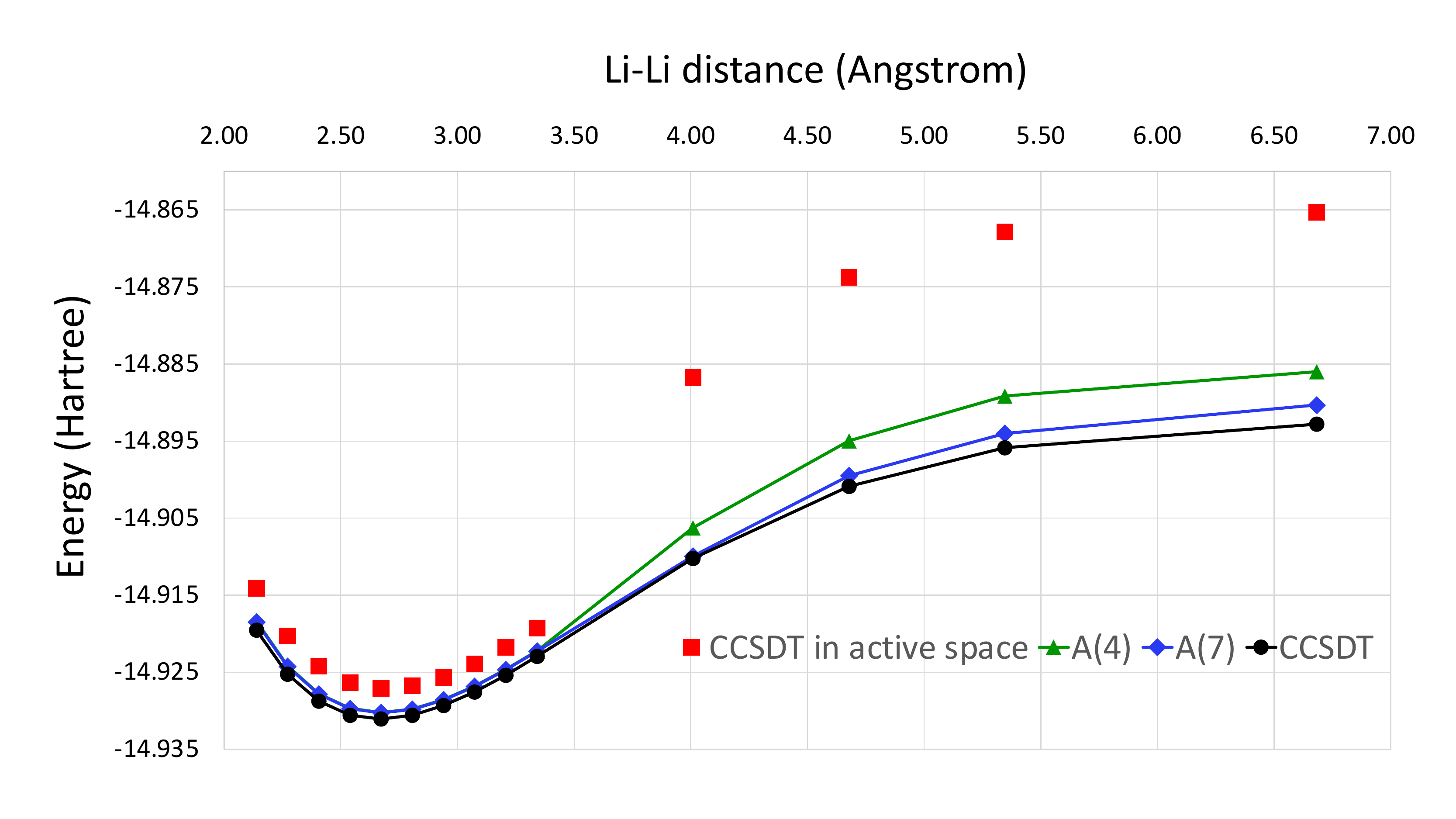}
	\caption{A comparison of the CCSDT in active space, A(4), A(7), and CCSDT ground-state potential energy surfaces for the Li$_2$ system in the cc-pVTZ basis set (see text for details). All calculations were performed with the MBPT(2) natural orbitals. The active spaces were defined by 10 MBPT(2) natural orbitals with  largest occupation numbers.}
\label{fig2}
\end{figure}

In this subsection, we will compare the performance of the A($i$) ($i=2,\ldots,7$) approximations for the Li$_2$ system in the cc-pVTZ basis set (spherical harmonic representation of $d$ orbitals was employed) containing 60 basis set functions. In analogy to previous studies\cite{metcalf2020resource} that demonstrated advantages of using correlated molecular orbital basis, we will use the natural orbitals calculated at the second order of M\o ller-Plesset perturbation theory (MBPT(2)). 
To define active spaces, we used 10 MBPT(2) natural orbitals characterized by the largest 
occupancies. 
The results of calculations are shown in Fig. \ref{fig2} and 
Tables \ref{table_li21} and \ref{table_li22}. In Fig. \ref{fig2} we show the potential energy surfaces corresponding to CCSDT within the active space, A(4), A(7), and CCSDT formulations (the values of calculated energies are shown
in Table \ref{table_li21}). It should be stressed that the full CCSDT formalism for Li$_2$ system  provides nearly FCI level of accuracy in the calculated ground-state energies. 
From Fig. \ref{fig2} one can see that energies obtained with the CCSDT approach in the active space
significantly underestimate dissociation energy (errors for large separations are significantly larger than the errors near the equilibrium geometry). One should notice that the A(7) formalism provides a very balanced description of correlation effects and provides further improvement of the A(4) results. This fact is well illustrated in Table \ref{table_li22}, where one can observe that the A(7) errors with respect to the full CCSDT energies do not exceed 2.5 milliHartree. The same error bound for A(4) formulations is larger and amounts to 6.9 milliHartree. 
It is also worth emphasizing that the high quality of the A(7) results was achieved despite the fact that the MBPT(2) formalism may not be the method of choice for large Li-Li separations and MBPT(2) natural orbitals may significantly differ from those defined by non-perturbative formalisms such as the CCSD approach.

\renewcommand{\tabcolsep}{0.2cm}
\begin{center}
  \begin{table*}
    \centering
    \caption{Total energies of Li$_2$ in the cc-pVTZ basis set (spherical representation of the $d$ orbitals was used)  obtained with various CC  formulations as functions of  Li-Li distance (in \r{A}). The A(4) and A(7) calculations  were performed using active spaces defined by 10 MBPT(2) natural  orbitals corresponding 
    to largest occupancies.}
    \begin{tabular}{lccccc} \hline \hline  \\
$R_{\rm Li-Li}$ & CCSD &  CCSDT & CCSDT       & A(4) & A(7) \\[-0.1cm]
                &      &        &  (active space)  &     &      \\[0.1cm]
    \hline \\
2.13840 & -14.91902 & -14.91953 & -14.91409 & -14.91836 & -14.91851 \\[0.0cm]
2.27205 & -14.92478 & -14.92524 & -14.92028 & -14.92417 & -14.92427 \\[0.0cm]
2.40570 & -14.92834 & -14.92876 & -14.92421 & -14.92778 & -14.92786 \\[0.0cm]
2.53935 & -14.93019 & -14.93057 & -14.92635 & -14.92967 & -14.92972 \\[0.0cm]
2.67300 & -14.93071 & -14.93107 & -14.92709 & -14.93023 & -14.93027 \\[0.0cm]
2.80665 & -14.93022 & -14.93057 & -14.92678 & -14.92978 & -14.92981 \\[0.0cm]
2.94030 & -14.92899 & -14.92933 & -14.92566 & -14.92857 & -14.92860 \\[0.0cm]
3.07395 & -14.92720 & -14.92754 & -14.92393 & -14.92681 & -14.92683 \\[0.0cm]
3.20760 & -14.92502 & -14.92536 & -14.92176 & -14.92465 & -14.92467 \\[0.0cm]
3.34125 & -14.92259 & -14.92294 & -14.91926 & -14.92223 & -14.92226 \\[0.0cm]
4.00950 & -14.90977 & -14.91024 & -14.88676 & -14.90624 & -14.90996 \\[0.0cm]
4.67775 & -14.90012 & -14.90085 & -14.87377 & -14.89497 & -14.89951 \\[0.0cm]
5.34600 & -14.89484 & -14.89586 & -14.86784 & -14.88918 & -14.89402 \\[0.0cm]
6.68250 & -14.89142 & -14.89281 & -14.86529 & -14.88600 & -14.89032  \\[0.2cm]
    \hline \hline 
    \end{tabular}
    \label{table_li21}
  \end{table*}
\end{center}
%
%
%
%
\renewcommand{\tabcolsep}{0.2cm}
\begin{center}
  \begin{table*}
    \centering
    \caption{Errors of the CCSDT in active space, A(4), and A(7) energies (in Hartree) with respect to full  CCSDT energies of the Li$_2$ described by the cc-pVTZ basis set
    (spherical representation of the $d$ orbitals was used) as functions of Li-Li distance (in \r{A}). 
    The A(4) and A(7) calculations  were performed using active spaces defined by 10 MBPT(2) natural  orbitals corresponding 
    to the largest occupancies.
     }
    \begin{tabular}{lccc} \hline \hline  \\
$R_{\rm Li-Li}$ & CCSDT &  A(4)  & A(7) \\[0.0cm]
 &     (active space) &     &  \\[0.1cm]
    \hline \\
2.13840 & 0.00544 &	0.00117	& 0.00103 \\[0.1cm]
2.27205	& 0.00495 &	0.00107	& 0.00096 \\[0.1cm]
2.40570	& 0.00455 &	0.00098	& 0.00090 \\[0.1cm]
2.53935	& 0.00423 &	0.00090	& 0.00085 \\[0.1cm]
2.67300	& 0.00398 &	0.00084	& 0.00080 \\[0.1cm]
2.80665	& 0.00380 &	0.00079	& 0.00076 \\[0.1cm]
2.94030	& 0.00367 &	0.00075	& 0.00073 \\[0.1cm]
3.07395	& 0.00360 &	0.00073	& 0.00071 \\[0.1cm]
3.20760	& 0.00360 &	0.00071	& 0.00069 \\[0.1cm]
3.34125	& 0.00368 &	0.00071	& 0.00068 \\[0.1cm]
4.00950	& 0.02348 &	0.00400	& 0.00028 \\[0.1cm]
4.67775	& 0.02708 &	0.00588	& 0.00134 \\[0.1cm]
5.34600	& 0.02802 &	0.00668	& 0.00184 \\[0.1cm]
6.68250	& 0.02752 &	0.00681	& 0.00249 \\[0.2cm]
    \hline \hline 
    \end{tabular}
    \label{table_li22}
  \end{table*}
\end{center}
%
%
%
%
%
%
\subsection{The H$_2$O molecule}
Breaking a single bond in the water molecule provides yet another example for illustrating the performance of the downfolding methods in describing chemical transformations. We also use this benchmark to analyze the convergence of the commutator expansions as a function of the active space size. 
In the present studies, we use the equilibrium geometry of the water molecule defined by the following parameters: $R_{OH}=R_e=0.96183$ \r{A}, $\angle_{HOH}=103.9215^{\circ}$. In Table \ref{table_h2o1} we compare the energies obtained with various downfolding methods (A(2)-A(7)) with CCSDTQ and CCSDTQ-in-active-space formulations for active space defined by 12 active RHF orbitals. 
The CCSDTQ-in-active-space helps calibrate the importance of the  correlation effects originating in the orthogonal complement of active space. For all geometries considered here, we used cc-pVTZ basis set (employing spherical representation of the $d$ orbitals), which yields 58 RHF orbitals. From Table \ref{table_h2o1} one can see that for all geometries the A(6) and A(7) approximations provide energies of similar values, which are consistently located above the CCSDTQ ones. For each geometry considered in Table \ref{table_h2o1} A(6) and A(7) provide significant improvements of the CCSDTQ-in-active-space energies. For example, the CCSDTQ-in-active-space 
errors with respect to the CCSDTQ energies where all orbitals are correlated amounts to 234, 230, and 227 milliHartree for $1.0R_e$, $1.5R_e$, and $2.0R_e$, respectively. For comparison, the analogous A(7) errors take values: 3, 10, and 22 milliHartree.
It should also be stressed that the perturbative A(5) approximation consistently underestimates the CCSDTQ energies by 5-7 milliHartree. The double commutator expansions A(5)-A(7) also improve the quality of single-commutator formulations (including the most complete A(4) formulation).

\begin{figure}
	\includegraphics[angle=0, width=0.48\textwidth]{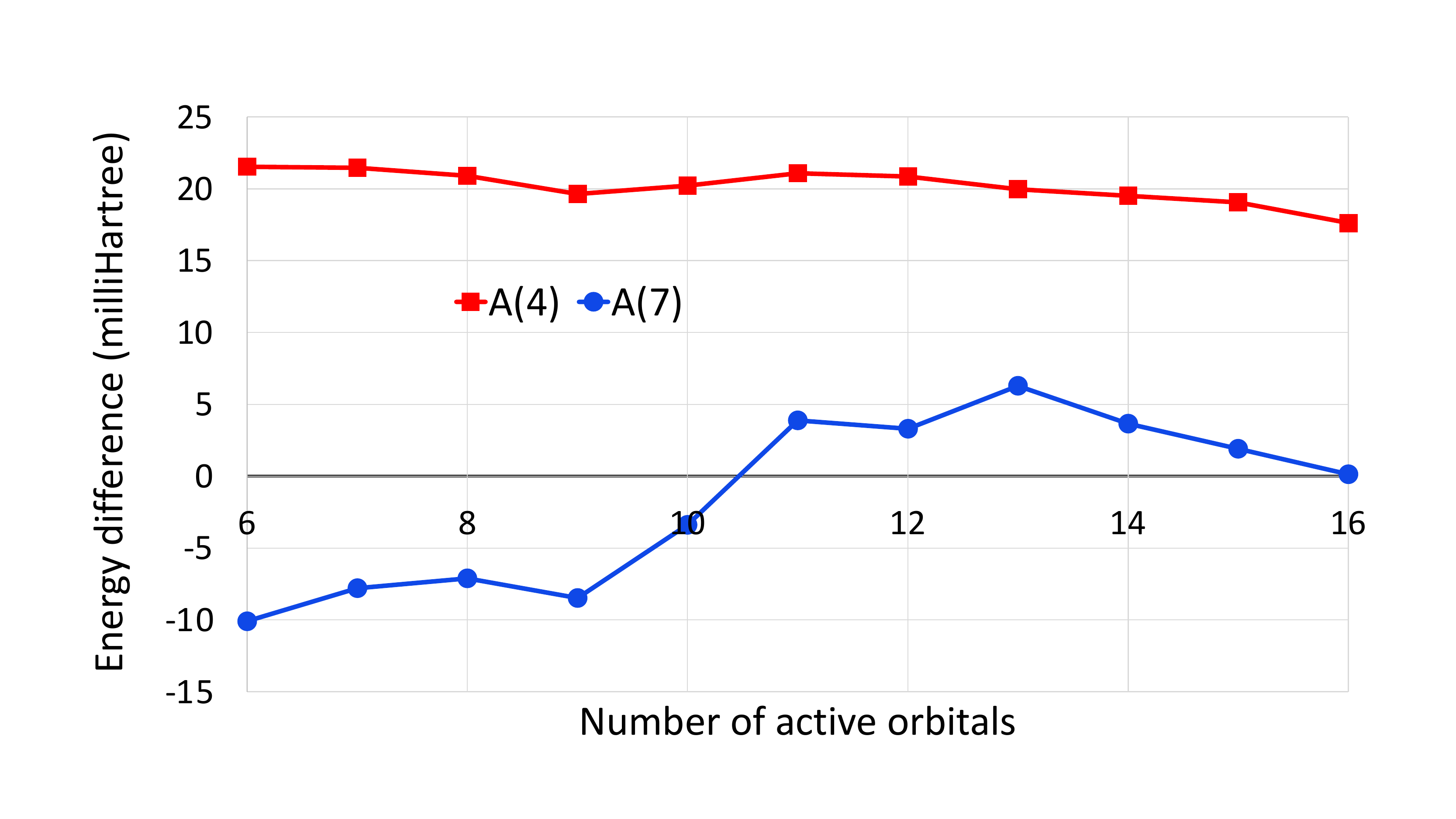}
	\caption{Errors of the A(4) and A(7) approximations as functions of the number of active orbitals with respect to the CCSDTQ energy (obtained by correlating all RHF molecular orbitals) for the equilibrium geometry of the H$_2$O system (see Table \ref{table_h2o1} and text for details) in the cc-pVTZ basis set (spherical representation of $d$ orbitals was employed). }
\label{fig3}
\end{figure}

To explore the impact of the size of the active space on the quality of downfolding formulations, we performed a series of calculations with the A(2)-A(7) approaches for the equilibrium structure of the water molecule for active spaces defined by 
$6,7,\ldots,16$ active orbitals corresponding to the lowest-lying RHF molecular orbitals (see Table \ref{table_h2o2}).
For the water structure where one OH bond is stretched to 2.0$R_e$ we perform similar studies with representative A(4) and A(7) approaches for active spaces defined by 12-16 active orbitals (see Table \ref{table_h2o3}).
It is justified to expect that the increase of the active space size can compensate for simplifications S1-S3 (discussed in Section II), which are indispensable in dealing with non-terminating series.  For example, part of the  effects due to external  triply ($T_{\rm ext,3}$) and quadruply $T_{\rm ext,4}$ excited amplitudes for small active spaces, which are not included in the discussed approximations, can be accommodated in the FCI-type solvers (quantum and classical) as internal excitations for larger active spaces. 
Indeed, by scrutinizing Table \ref{table_h2o2} and Fig. \ref{fig3} one can observe that for larger active space comprising of 16 orbitals (which still provides a nearly four-fold reduction of the total size of orbital space), the A(7) results provide nearly CCSDTQ level of accuracy. 
At the same time, A(7) results are more accurate than the energies obtained with single commutator expansion A(4) (see Fig. \ref{fig3}) and converge quickly to the CCSDTQ level of accuracy. This is happening despite the fact that a very simple form of external $T_{\rm ext}$ is used and that the CCSDTQ-in-active-space energy for active space defined by 16 orbitals  is characterized by 193 milliHartree of error 
with respect to the full CCSDTQ energy. 

In Table \ref{table_h2o3} we can observe a similar behavior for 2.0$R_e$ case. In this situation, the error of the A(7) energies with respect to the CCSDTQ data for 16 orbital active space is characterized by a larger error (6.746 milliHartree) compared to the equilibrium case. However, given the reduction of 184.505 milliHartree error obtained with the CCSDTQ-in-active-space approach, one should consider A(7) accuracy as quite satisfying given the number of simplifications (S1-S3) used to define the corresponding A(7) downfolded Hamiltonian. It is also instructive to observe that the quality of the 16-active orbital A(7) energy is significantly better than the one obtained with the full CCSD approach (17.839 milliHartree of error), which provides the source for the external amplitudes. 
\renewcommand{\tabcolsep}{0.2cm}
\begin{center}
  \begin{table*}
    \centering
    \caption{A comparison of the A$(i)$ $i=2,\ldots7$ energies with those calculated with CCSDTQ in active space 
     and CCSDTQ formalisms obtained for single bond breaking for water molecule in cc-pVTZ basis set (with spherical representation of the $d$ functions). The parameters defining the equilibrium geometry of the water molecule are defined as $R_{OH}=R_e=0.96183$ \r{A}, $\angle_{HOH}=103.9215^{\circ}$. 
     For all geometries  active spaces were defined by 12 lowest-lying RHF molecular orbitals.}
    \begin{tabular}{lccc} \hline \hline  \\
  Method &  1.0R$_e$ &  1.5R$_e$ & 2.0R$_e$  \\
\hline \\
CCSDTQ  & -76.34615 &  -76.25808 & -76.18435    \\[0.2cm]
CCSDTQ  &-76.11177  &  -76.02825 & -75.95730    \\[-0.1cm]
(active space) &   &    &                            \\[0.1cm]
A(2) &	-76.32581	&-76.23560&	-76.15386 \\[0.1cm]
A(3) &	-76.56782	&-76.47198&	-76.38497 \\[0.1cm]
A(4) &	-76.32530	&-76.23516&	-76.15456 \\[0.1cm]
A(5) &	-76.35205	&-76.26556&	-76.18986 \\[0.1cm]
A(6) &	-76.34305	&-76.24817&	-76.16284 \\[0.1cm]
A(7) &	-76.34286	&-76.24801&	-76.16270 \\[0.2cm]
    \hline \hline 
    \end{tabular}
    \label{table_h2o1}
  \end{table*}
\end{center}
\renewcommand{\tabcolsep}{0.05cm}
\begin{center}
  \begin{table*}
    \centering
    \caption{Energies of H$_2$O for equilibrium geometry ($R_{OH}=R_e=0.96183$ \r{A}, $\angle_{HOH}=103.9215^{\circ}$ ) obtained with the  cc-pVTZ basis set (with spherical representation of $d$ orbitals) and various sizes of active spaces (from 6 to 16 active orbitals; "all orbitals" means that the CCSD, CCSDT, and CCSDTQ calculations were performed correlating all RHF molecular orbitals).
     }
    \begin{tabular}{lcccccccccccc} \hline \hline  \\
  Method &  6 & 7 & 8 & 9 & 10 & 11 & 12 & 13 & 14& 15 & 16 & all orbitals    \\
\hline \\
CCSD &	-76.05760 & -76.05977&	-76.06491&	-76.07286&	-76.08267&	-76.10487&	-76.11028&	-76.12626&	-76.13004&	-76.13379&	-76.15033&	-76.33793 \\[0.2cm]
CCSDT &	-76.05760&	-76.05982&	-76.06500&	-76.07301&	-76.08316&	-76.10634&	-76.11195&	-76.12825&	-76.13236&	-76.13641&	-76.15338&	-76.34582 \\[0.2cm]
CCSDTQ& 	-76.05760&	-76.05982&	-76.06500&	-76.07300&	-76.08313&	-76.10617&	-76.11177&	-76.12806&	-76.13216&	-76.13619&	-76.15312&	-76.34615 \\[0.2cm]
A(2)&	-76.32635&	-76.32637&	-76.32684&	-76.32802&	-76.32716&	-76.32566&	-76.32581&	-76.32633&	-76.32674&	-76.32718&	-76.32838\\[0.2cm]	
A(3)&	-76.61835&	-76.61627&	-76.61123&	-76.60349&	-76.59423&	-76.57301&	-76.56782&	-76.55225&	-76.54879&	-76.54529&	-76.52932	&   \\[0.2cm]
A(4)&	-76.32462&	-76.32469&	-76.32526&	-76.32652&	-76.32595&	-76.32507&	-76.32530&	-76.32619&	-76.32664&	-76.32711&	-76.32853&  \\[0.2cm]	
A(5)&	-76.35393&	-76.35286&	-76.35278&	-76.35452&	-76.35165&	-76.35084&	-76.35205&	-76.34770&	-76.35036&	-76.35248&	-76.35497& \\[0.2cm]	
A(6)&	-76.35622&	-76.35395&	-76.35332&	-76.35470&	-76.34971&	-76.34246&	-76.34305&	-76.34005&	-76.34268&	-76.34442&	-76.34612& \\[0.2cm]	
A(7)&	-76.35624&	-76.35394&	-76.35327&	-76.35462&	-76.34953&	-76.34226&	-76.34286&	-76.33987&	-76.34251&	-76.34425&	-76.34603& \\[0.2cm]	
    \hline \hline 
    \end{tabular}
    \label{table_h2o2}
  \end{table*}
\end{center}
\renewcommand{\tabcolsep}{0.05cm}
\begin{center}
  \begin{table*}
    \centering
    \caption{Energies of H$_2$O corresponding to geometry where a single OH bond is stretched to twice the equilibrium distance, denoted $2.0R_e$  ($R_{OH}=R_e=0.96183$ \r{A}, $\angle_{HOH}=103.9215^{\circ}$ ). Results were obtained in the cc-pVTZ basis set (with spherical representation of $d$ orbitals) with various sizes of active spaces (from 12 to 16 active orbitals; "all orbitals" means that the CCSD, CCSDT, and CCSDTQ calculations were performed correlating all RHF molecular orbitals).
     }
    \begin{tabular}{lcccccc} \hline \hline  \\
  Method &   12 & 13 & 14& 15 & 16 & all orbitals    \\
\hline \\
CCSD  &	-75.95533 & -75.96964 & -75.98227 & -75.99151 &	-75.99468 & -76.16651 \\[0.2cm]
CCSDT &	-75.95732 & -75.97217 & -75.98566 & -75.99584 &	-75.99920 & -76.18293 \\[0.2cm]
CCSDTQ& -75.95730 & -75.97236 & -75.98613 & -75.99645 &	-75.99984 & -76.18435 \\[0.2cm]	
A(4)  &	-76.15456 & -76.15494 & -76.15777 & -76.15884 &	-76.15918 &            \\[0.2cm]
A(7)  &	-76.16270 & -76.15865 & -76.17306 & -76.17523 &	-76.17760 &            \\[0.2cm]
    \hline \hline 
    \end{tabular}
    \label{table_h2o3}
  \end{table*}
\end{center}
%
%
%
%
%
%
%
%
%
\section{Conclusions}
In this paper, we discuss the level of accuracy provided by the active-space downfolded Hamiltonians, including variants involving non-trivial terms stemming from double commutators. In analogy to previous studies, for the sake of perturbative consistency, in some approximations, we also included Fock-operator-dependent terms stemming from triple commutators. 
In all approximations analyzed in this paper, we used a subset of  CCSD cluster amplitudes to 
evaluate the so-called external part of the anti-Hermitian cluster operator $\sigma_{\rm ext}$. 
To assess the performance of these approximations we performed a series of DUCC simulations for typical benchmark systems such as the beryllium atom, Li$_2$ dimer, and the water molecule. Using these examples, we demonstrated that the inclusion 
of double-commutator terms results in consistent improvements of the DUCC approaches based on a single-commutator expansion. This was the best seen on the example of a bond breaking in the Li$_2$ system when the integration of the A(7) approach with natural orbitals for modest size active space resulted in nearly CCSDT level of accuracy for all geometries considered. 
Using H$_2$O example, we also demonstrated that the effect of S1-S3 approximations needed to deal with non-terminating series can be, to a large extent, accommodated by enlarging the size of the active space. At the same time, we demonstrated that A(7) energies quickly approach the full CCSDTQ results as the size of the active space increases. The double-commutator-based approximation A(7) not only outperforms the single-commutator-based  A(4) approach but can also provide CCSDTQ level  quality for active spaces which offer nearly four-fold reduction in the size of orbital space employed. 

A natural extension of the current studies is to include  higher-many-body components in the downfolded Hamiltonians. In particular, in the forthcoming papers we will include: (1)  3-body interactions (in the particle-hole representation) stemming from contractions of two-body part of the electronic Hamiltonian with $\sigma_{\rm ext,2}$ operators and (2) terms due to the $\sigma_{\rm ext,3}$ and/or $\sigma_{\rm ext,4}$ operators. 

We believe that energy accuracies obtainable with downfolding techniques are getting us close to quantum simulations of realistic chemical processes. The downfolding Hamiltonians also open an opportunity for utilizing machine learning methods to extract the analytical form of the effective inter-electron interactions in small-dimensionality sub-spaces of the entire Hilbert space.

%

\section{acknowledgement}
This work was supported by the "Embedding QC into Many-body Frameworks for Strongly Correlated  Molecular and Materials Systems'' project, which is funded by the U.S. Department of Energy, Office of Science, Office of Basic Energy Sciences (BES), the Division of Chemical Sciences, Geosciences, and Biosciences.
Part of this work was supported by  the Quantum Science Center (QSC), a National Quantum Information Science Research Center of the U.S. Department of Energy (DOE).

\section{data availability}
The data that support the findings of this study are available from the corresponding author upon reasonable request.




%

\end{document}